\documentclass[%
 reprint,
 amsmath,amssymb,
 aps,
]{revtex4-2}

\usepackage{graphicx}
\usepackage{dcolumn}
\usepackage{bm}

\begin{document}

\preprint{APS/123-QED}

\title{Simultaneous High-Efficiency Anomalous Reflection and Angle of Arrival Sensing in Reconfigurable Intelligent Surfaces
}

\author{Mostafa Movahediqomi$^{1}$}
 \email{mostafa.movahediqomi@aalto.fi}
\author{Yongming Li$^{1,2}$}


\author{Grigorii Ptitcyn$^{3}$}
\author{Viktar Asadchy$^{1}$}

\author{Sergei Tretyakov$^{1}$}
\affiliation{%
$^{1}$Department of Electronics and Nanoengineering, Aalto University, P.O. Box 15500, FI-00076 Aalto, Finland\\
$^{2}$State Key Laboratory of Electrical Insulation and
Power Equipment, School of Electrical Engineering, \\Xi'an Jiaotong University, Xi'an 710049, China\\
$^3$University of Pennsylvania, Department of Electrical and Systems Engineering, Philadelphia, PA 19104, U.S.A.
}

\date{\today}

\begin{abstract}

In this work, we introduce reconfigurable intelligent surfaces designed to simultaneously perform reflection of single or multiple incident waves toward the receiver or receivers and sensing the angles of arrival.  
We achieve anomalous reflection with strongly suppressed parasitic scattering through an in-situ optimization of either the currents flowing on array elements or the far field in the receiver direction. The suppression of parasitic scattering allows us to accurately and without additional measurements or computations detect the angles of arrival of the illuminations through the spatial Fourier transform of the optimized current distribution through the controllable reactive loads. Therefore, unlike other  recently proposed methods, our scheme of integrated sensing and communication does not require any pre-computed data sets and works for an arbitrary number of simultaneous illuminations. As a proof of principle, we design and analyze with full-wave simulations several reconfigurable intelligent surfaces consisting of an array of loaded wires above a ground plane.
%
\end{abstract}

\maketitle



\section{Introduction}

In recent developments of wireless communications, multifunctional metasurfaces have attracted substantial attention due to their remarkable capacity to execute multiple tasks within a unified platform~\cite{wang2020independent,asadchy2017flat,yin2019terahertz,zhang2017multichannel,zhang2019multichannel,kamali2017angle,wang2020theory,chen2020directional,malek2022multifunctional,overvig2022diffractive}. Among this category, integrated sensing and communication (ISAC) metasurfaces stand out as promising solutions for the upcoming generation of mobile networks, specifically 6G. These metasurfaces exhibit the ability to perform anomalous and tunable reflection and simultaneously engage in sensing of propagation environments to estimate the angles of arrival (AoA). Early investigations revealed the potential of antennas as tools for AoA estimation using Fourier techniques, although these efforts primarily centered around channel analysis~\cite{kalliola2000real}. More recently, studies have spotlighted reconfigurable intelligent surfaces (RISs) as a means of some limited-functionality sensing. In papers~\cite{wang2020channel,zheng2021efficient}, RISs that transmit pilot sequences between base stations (BS) and user equipment (UE) have been proposed. However, a significant challenge arises from potential failures to properly decouple transmitter and receiver RISs. An alternative solution involves augmenting the system with an extra sensing antenna and RF chain~\cite{liu2020deep,taha2021enabling}. However, that approach comes with the trade-offs of diminishing the effective coverage area of reflective cells and necessitating an additional expensive RF chain. An interesting proposition surfaced in~\cite{liaskos2019absense}, advocating the absorption of incoming electromagnetic waves through an RIS embedded with a sensor system. This system gauges the absorbed energy against a prepared lookup bank, although its practicality is constrained by the limited resolution of available datasets and the absence of experimental validation. Recently, a novel sensing paradigm arose wherein a portion of the total wave power is employed for sensing, being directed toward a sensing waveguide positioned beneath the RIS~\cite{alexandropoulos2023hybrid,zhang2023channel,zhang2021channel,alamzadeh2023detecting,alamzadeh2021reconfigurable,alamzadeh2022sensing,albanese2022marisa,albanese2023ares,jiang2023simultaneously}. However, an inherent drawback lies in this approach: the sensing process takes place across the entire fields at the RIS interface, encompassing both incident and scattered fields. The challenge emerges from the unknown direction of incident waves, leading to complexities in segregating these fields. Consequently, multiple pre-computed test RIS illuminations are necessary. Furthermore, this approach is constrained to single plane-wave illumination due to the waveguide geometry. Additionally, post-sampling RF combiners are required to aggregate signals before transmission to RF detectors, contributing to escalated complexity and costs associated with this method. Another work was done to merge sensing of the direction of arrival with radar cross-section (RCS) manipulation by using space-time modulated metasurfaces~\cite{fang2023multifunctional}. Estimation was performed in the transmission region (behind the metasurface), and the RCS manipulation was done in the reflection region (in front of the metasurface).

\begin{figure*}
     \centering
    \includegraphics[width=1\textwidth]{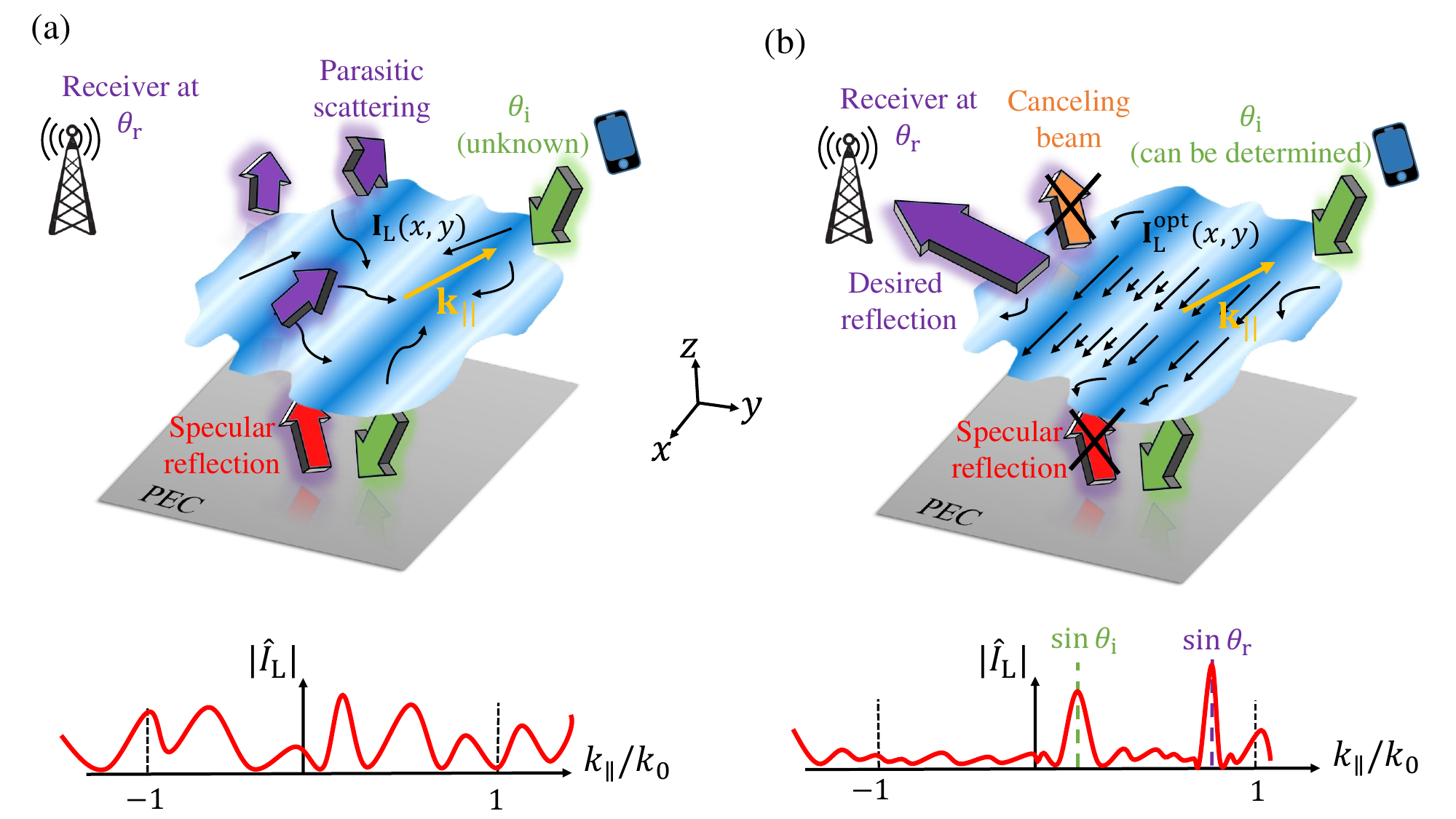}
    \caption{An illustration of the principle of the proposed solution. (a) Before optimization, a random distribution of load impedances leads to a complicated reflected field. The bottom plot qualitatively illustrates the Fourier transform of surface-averaged electric current passing through the loads along a certain direction (in the particular examples below that will be in the incidence plane, ${\bf k}_{||}=k_y{\hat y}$). (b) By optimizing the load impedances and subsequently the current distribution in the controllable array elements, the RIS anomalously reflects a single plane wave to the receiver direction. At the bottom, the spatial Fourier transform of the optimized current is shown. Because the optimal anomalous reflection implies elimination of the specular reflection from the ground plane, the second peak position reveals the direction of arrival of the incident wave.}
    \label{fig:conceptual}
\end{figure*}

This paper presents an alternative approach for realizing ISAC functionalities in RISs. Unlike the known methods, the proposed technique only requires knowledge of the incident wave's frequency. This feature stands in stark contrast to existing approaches that rely on pre-computed data sets. The method is based on in-situ optimization of reflection toward the receiver device that does not require knowledge of the incident wave characteristics (e.g., polarization and AoA). We showcase two types of optimization, based on the currents flowing on the RIS elements and based on the far field strength at the receiver. After the optimization, the impedance mismatch between the incoming and reflected waves toward the receiver direction is minimized. We demonstrate that this approach can effectively achieve dual objectives in the same platform: nearly ideal power reflection toward the intended receiver position as well as efficient AoA sensing of the incident waves. This concept is based on the fact that optimization of the current distribution for the desired reflection results in the creation of a phase-synchronized component of voltages and currents at the RIS elements that aligns with the incident wave space variations. This component compensates for the specular reflection from the RIS's ground plane, or, in more general cases, the structural scattering of the antenna array. Therefore, through a spatial Fourier transform of the optimized current distribution we can determine the direction of the incident wave propagation.  
We demonstrate our method in scenarios involving single and simultaneous multiple illuminations from various directions.

\section{Concept description}

The general concept of the proposed RIS supporting ISAC is illustrated in~Fig.~\ref{fig:conceptual}. The device consists of some tunable scattering structure shown in the figure with blue color (for example, an array of small metal patches with controllable loads) positioned above a reflecting ground plane. 
Reflection can be optimized by optimizing the distribution of current $I_L(x,y)$ induced on the scattering structure (for example, varying reactive loads of the array patches).  It is assumed that the receiver position, i.e., the angle $\theta_\mathrm{r}$, is \textit{known}. However, the direction of the incoming wave (defined by the angle of incidence $\theta_\mathrm{i}$ relative to the surface normal direction or the $z$-axis) is \textit{unknown}. The goal is to maximize the signal reflected to the direction toward the receiver and simultaneously find the angle of incidence. 
We do not assume periodic current distribution along the array, and all the load impedances of array elements can be different. 
In the last section, we will discuss also the more general case when the array is illuminated by plane waves from several unknown directions at the same time. 

To elucidate the proposed concept, 
let us first consider the scenario when the array initially has a random load impedance distribution. When illuminated by an incident wave at an unknown angle $\theta_{\rm i}$, electric currents will be excited in the structure. The current distribution on the plane will be, in general, non-uniform, denoted as $I_{\rm L}(x,y)$. 
The reflected field from the RIS consists mostly of parasitic scattering and specular reflection from the ground plane. Only a small portion of the reflected power is delivered to the receiver in such random configuration. This situation is illustrated in Fig.~\ref{fig:conceptual}(a).

The distribution of the induced current is convenient to characterize by its 
spatial Fourier transform
\begin{equation}
\label{Eq:transform}
\widehat{\bf I}_{\rm L} ({\bf k}_{||}) = \int_{-\infty}^{\infty} \mathrm {\bf I}_{\mathrm L}(x,y) e^{-j{\bf k}_{||}\cdot {\bm \rho}} \, {\rm d}x{\rm d}y, 
\end{equation}
where ${\bm \rho}=x\hat x+y\hat y$ and ${\bf k}_{||}=k_x\hat x+k_y\hat y$. For electrically large reflectors, the amplitudes of waves scattered into directions $\sin\theta= |{k}_{||}|/k_0$ are defined by the amplitudes of the corresponding spectral components of the current distribution. Here, $k_0$ is the free-space wavenumber.

As the next step, we optimize the current distribution (by varying the load impedances $Z_{\rm L}$ of the array elements) in order to maximize the power reflected toward the receiver. This optimization is \textit{in-situ}, i.e., it is made in real time, being updated if the illumination changes. Importantly, the optimization does not require knowledge of the incident wave AoA and only needs the receiver's angle $\theta_{\rm r}$. In Section \ref{secoptim}, two different types of optimization are considered. Once the in-situ optimization has converged, the RIS creates a dominant \textit{single} beam in the direction of the receiver. Parasitic reflections in the other directions are automatically minimized due to the conservation of energy, when reflection toward the desired direction is maximized.  This scenario is depicted in Fig.~\ref{fig:conceptual}(b). 

Now, we are ready to determine the AoA of the incident plane wave. 
First, we measure the final (after the optimization) values of the electric current $I_L$ flowing on the plane. Then using  (\ref{Eq:transform}), we calculate the spatial current spectrum. By plotting the spectrum, we will see only two dominant peaks, as shown at the bottom of Fig.~\ref{fig:conceptual}(b). The first current peak (shown in purple) corresponds to the generation of the anomalously reflected beam in the desired direction at $k_{||}/k_0 = \sin \theta_{\rm r}$. The second peak (shown in green) reveals that the scattering structure creates an auxiliary beam in some other direction. However, as we mentioned above, the optimization must result in only one reflected  beam at $\theta_{\rm r}$. This implies that the auxiliary beam radiated by the scattering structure  
can be nothing else than the beam that partially cancels (via destructive interference) the specular reflection generated by the ground plane itself. Since the specular reflection from the ground plane must be at the same angle as the incident angle $\theta_{\rm i}$, we conclude that the location of the green peak $k_y/k_0$ in Fig.~\ref{fig:conceptual}(b) must be equal to $\sin \theta_{\rm i}$. Thus, from the measured data of the induced currents in the optimized RIS, we can determine the unknown angle of arrival $\theta_{\rm i}$.

\section{Model geometry and theory}

\begin{figure}[tb]
     \centering
         \includegraphics[width=0.5\textwidth]{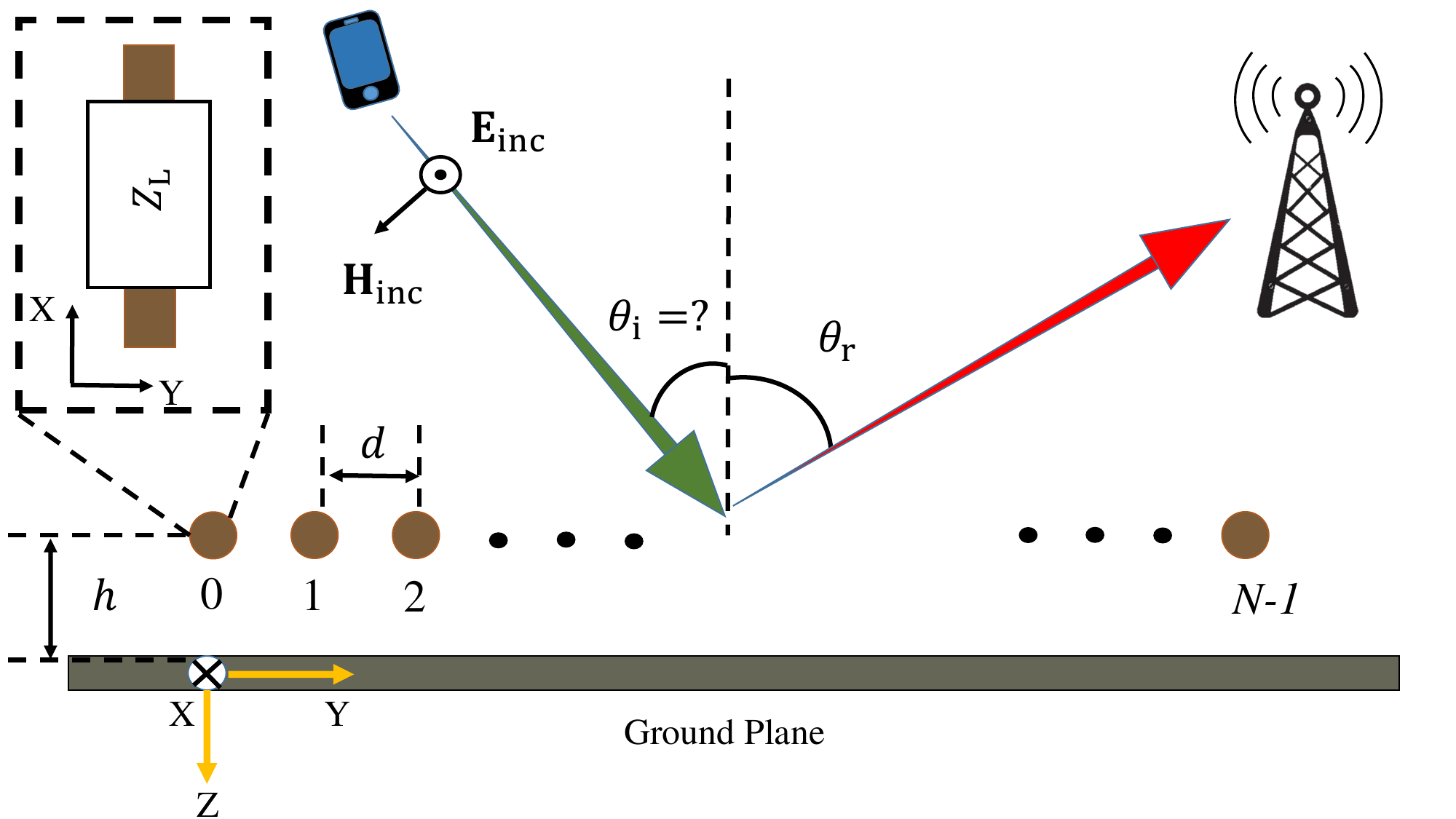}
    \caption{An illustration of an array of loaded wires above a PEC ground plane. The structure is illuminated with a TE polarized plane wave at an unknown angle $\theta_{\rm i}$. The wires are oriented horizontally along the $y$-axis at the distance $h$ from the ground plane and separated by $d$ from each other. The inset shows the $xy$-plane view of a single wire loaded by a tunable distributed impedance. The final goal is to create a reflected beam toward the receiver direction by optimizing the tunable loads and determine the incident angle $\theta_{\rm i}$.}
    \label{fig:array_wires}
\end{figure}

As a simple model platform suitable for analytical calculations, here we consider a two-dimensional model of finite-sized array of infinite wires over an infinitely large perfectly conducting ground plane. The wires are periodically loaded with load impedances $Z_{\rm L}$, defined as the load impedance per unit length of the wires. As we target to full power reflection, it is assumed that the loads are purely reactive. In practice, the loads can be realized by tunable capacitors, e.g., varactors, PIN diodes, or MEMS capacitors. 
As depicted in~Fig.~\ref{fig:array_wires}, the arrangement involves an array of $N$ wires aligned along the $x$-axis and arranged periodically along the $y$-axis with the period $d$.  The array is positioned above the ground plane at a distance $h$, with each wire having a radius of $r_0$. The material of the wires is assumed to be PEC. 
Because the currents flow in negligibly thin wires, the current distribution can be represented as a sum of Dirac delta functions, corresponding to the positions shown in~Fig.~\ref{fig:array_wires}. That is,  $I_L(y)=I_0\delta(y)+I_1\delta(y-d)+...+I_{N-1}\delta[y-(N-1)d]$. By substituting it to~Eq.~\eqref{Eq:transform}, the final expression can be achieved by a trivial calculation of the integral of delta functions:
\begin{equation}
\label{Eq:spatial_spect}
\widehat{I}_{\rm L} (k_y) = \sum_{m=0}^{N-1} I_m e^{-jk_y y_m}.
\end{equation}
Here, $I_m$ is the amplitude of the currents in each wire, and $y_m=md$ are their coordinates along $y$. 

In the pursuit of achieving reflection toward arbitrary directions, the traditional approach has been rooted in the phased-array principle, often referred to as the generalized law of reflection~\cite{huang2007reflectarray,yu2011light}. However, in such scenarios, a notable portion of the energy can scatter into other directions, resulting in a reduction of the power efficiency~\cite{movahediqomi2023comparison}. A possible solution to address this deficiency involves the utilization of nonlocal metasurfaces and metagratings~\cite{asadchy2016perfect,wang2020independent,kwon2018lossless,budhu2020perfectly,kwon2021planar,yepes2021perfect,diaz2017generalized,wong2018perfect,vuyyuru2023efficient,ra2017metagratings,ra2018reconfigurable,epstein2017unveiling,rabinovich2018analytical}. More recently, optimization solutions for aperiodic structures have emerged, offering the distinct advantage of unrestricted optimization due to the absence of periodicity constraints on current distribution at the reflector~\cite{li2023all,vuyyuru2023efficient2}. However, the optimization process employed in \cite{li2023all} required knowledge of the incident angle. Here, we propose an optimization strategy centered around load reactances as the optimization parameters to search for a global solution without any constraints. 
This method requires measurements of voltages or currents induced in the loads of the individual array elements for specific sets of reactive loads connected to the antenna ports. In this work, we emulate these measurements by analytical calculations of the induced currents. 

As an optimization method we employ the \textit{fmincon} optimization algorithm within the MATLAB environment. The foundational matrix equation for our design adheres to Ohm's law for currents induced in the wires, expressed as follows:
\begin{equation}
\label{Eq:ohm's_law}
\overline{\overline{Z}}_\mathrm{tot} \cdot \vec{I}_\mathrm{L}= {\vec  E}_\mathrm{tot}.
\end{equation}
Here, vector $\vec{I}_{\rm L}$ with $N$ elements includes the complex amplitudes of currents in the wires, $N$-elements vector $\vec{E}_\mathrm{tot}$ represents the total external field at the locations of the wires (the sum of the incident wave or waves and their reflections from the ground plane). 

In Eq.~(\ref{Eq:ohm's_law}),  $\overline{\overline{Z}}_\mathrm{tot}$ is the $N\times N$  impedance matrix of the array. This matrix comprises three distinct constituents: $\overline{\overline{Z}}_\mathrm{in}$, $\overline{\overline{Z}}_\mathrm{L}$, and $\overline{\overline{Z}}_\mathrm{mut}$, representing the self-impedance, load impedance, and mutual impedance matrices, respectively. Because the wires differ only by their loads, the input impedance matrix $\overline{\overline{Z}}_\mathrm{in}$ is a unit matrix multiplied by a scalar $Z_{\rm in}$:
\begin{align}
\overline{\overline{Z}}_\mathrm{in} &= Z_{\rm in}\overline{\overline{I}} .
\label{Eq:definition1}
\end{align}
For our geometry we get~\cite{li2023all,popov2018controlling}
\begin{align}
Z_{\rm in}&= \frac{k_0\eta}{4} \left[ H_0^{(2)}(k_0 r_0) - H_0^{(2)}(2 k_0 h)\right]. \label{Eq:definition2}
\end{align}
Here, 
$\eta = \sqrt{\mu_0 / \epsilon_0}$ is the free-space wave impedance, and function $H_0^{(2)}(\cdot)$ denotes the zeroth-order Hankel function of the second kind.

The load impedance matrix $\overline{\overline{Z}}_\mathrm{L}$ is also diagonal, although each load can be tuned independently, meaning that there are different values on the diagonal.  
Following the same logic one can get an expression for mutual impedances
\begin{align}
\overline{\overline{Z}}_\mathrm{mut} &= \begin{pmatrix}
    0 & Z_{0,1} & Z_{0,2} & \dots & Z_{0,N-1} \\
    Z_{1,0} & 0 & Z_{2,0} & \dots & Z_{1,N-1} \\
    Z_{2,0} & Z_{2,1} & 0 & \dots & Z_{2,N-1} \\
    \vdots & \vdots & \vdots & \ddots & \vdots \\
    Z_{N-1,0} & Z_{N-1,1} & Z_{N-1,2} & \dots & 0
\end{pmatrix},
\end{align}
where 
\begin{align}
    Z_{p,q} = -\frac{k_0\eta}{4} \Big[ H_0^{(2)}(k_0 |q-p|d) - \nonumber\\H_0^{(2)}(k_0 \sqrt{(|q-p|d)^2 + (2h)^2}) \Big].
\end{align}
The subscripts $q$ and $p$ are indices denoting the specific wire numbers for which the mutual impedance is being computed.

Looking at~Eq.~(\ref{Eq:ohm's_law}), it 
appears that in order to solve it for the currents, it is necessary to know the incident field. However, in practice the optimization process can be based solely on the measured values of currents induced in the wires. Here, we use this equation only to \textit{emulate} these currents measurements. In fact, from the known currents and the impedance matrix, the external field can be in principle found. 

As delineated during the formulation of~Eq.~(\ref{Eq:ohm's_law}) at the outset, we opt to consider an example of a TE-polarized wave incident from an arbitrary direction $\theta_\mathrm{i}$. In this case, at the locations of the wires ($y_{n}=nd$), electric field $\vec{E}_\mathrm{tot}(y,z)$ 
takes the form
\begin{equation}
\begin{aligned}
{\vec E}_\mathrm{tot}(y_n,-h) = &{\vec E}_\mathrm{inc}+{\vec E}_\mathrm{ref}\\
= &\begin{aligned}[t]
j 2 E_0 \sin (k_0 \cos \theta_{\rm i} h) 
\begin{bmatrix}
e^{- j k_0 \sin\theta_{\rm i} (0) d}\\
e^{- j k_0 \sin\theta_{\rm i} (1) d}\\
\vdots\\
e^{- j k_0 \sin\theta_{\rm i} (n) d}
\end{bmatrix}.
\end{aligned}
\end{aligned}
\label{Eq:total_field}
\end{equation}
Within every iteration of the optimization process, we update the load impedance values and compute the induced current vector using 
\begin{equation}
\label{Eq:current_cal}
\vec{I}_\mathrm{L}= \overline{\overline{Z}}_\mathrm{tot}^{-1} \cdot  {\vec E}_\mathrm{tot},
\end{equation}
where $\overline{\overline{Z}}_\mathrm{tot}^{-1}$ is the inverse of the total impedance matrix introduced earlier.
The only constraint imposed on the allowed loads during the optimization process is that they should be purely reactive. 
The initial values of the loads can be assigned arbitrarily. We stress once again that in this paper, Eq.~(\ref{Eq:current_cal}) is used only for emulating the measurements of currents or voltages. In a real implementation, the currents $\vec{I}_\mathrm{L}$ are measured directly across the loads, and Eq.~(\ref{Eq:current_cal}) is not used.

For the optimization problem, a variety of objective functions are available for selection. In this study, we adopt two distinct types of objective functions and subsequently present the outcomes associated with each case.

\section{Optimization methods and results}\label{secoptim}

\subsection{Optimization based on the current flowing on the array elements}

Assuming that the direction toward the receiver is known ($\theta_{\rm r}$), one potential approach involves maximizing the spatial frequency harmonic of the current distribution over the array  $\widehat{I}_{\rm {L}}(k_{||})={\rm F}\left\{ I_{\rm {L}} (x)\right\}$ that corresponds to the wave reflected toward $\theta_{\rm r}$. Here, F denotes the Fourier transform operation \eqref{Eq:transform} or \eqref{Eq:spatial_spect}, $k_{||}$ is the tangential wavenumber or the spatial frequency, and the target harmonic of the current corresponds to $\widehat{I}_{\rm {L}}(k_0 \sin \theta_{\rm {r}})$. For an infinite array, this particular spatial harmonic generates a reflected plane wave at the desired angle $\theta_{\rm {r}}$. Significantly, given the lossless nature of all load impedances, the maximization of this specific current harmonic within the spatial spectrum inherently implies optimization of the RIS efficiency, regardless of the unknown incident angle $\theta_\mathrm{i}$. 

\begin{figure}[tb]
     \centering
         \includegraphics[width=0.5\textwidth]{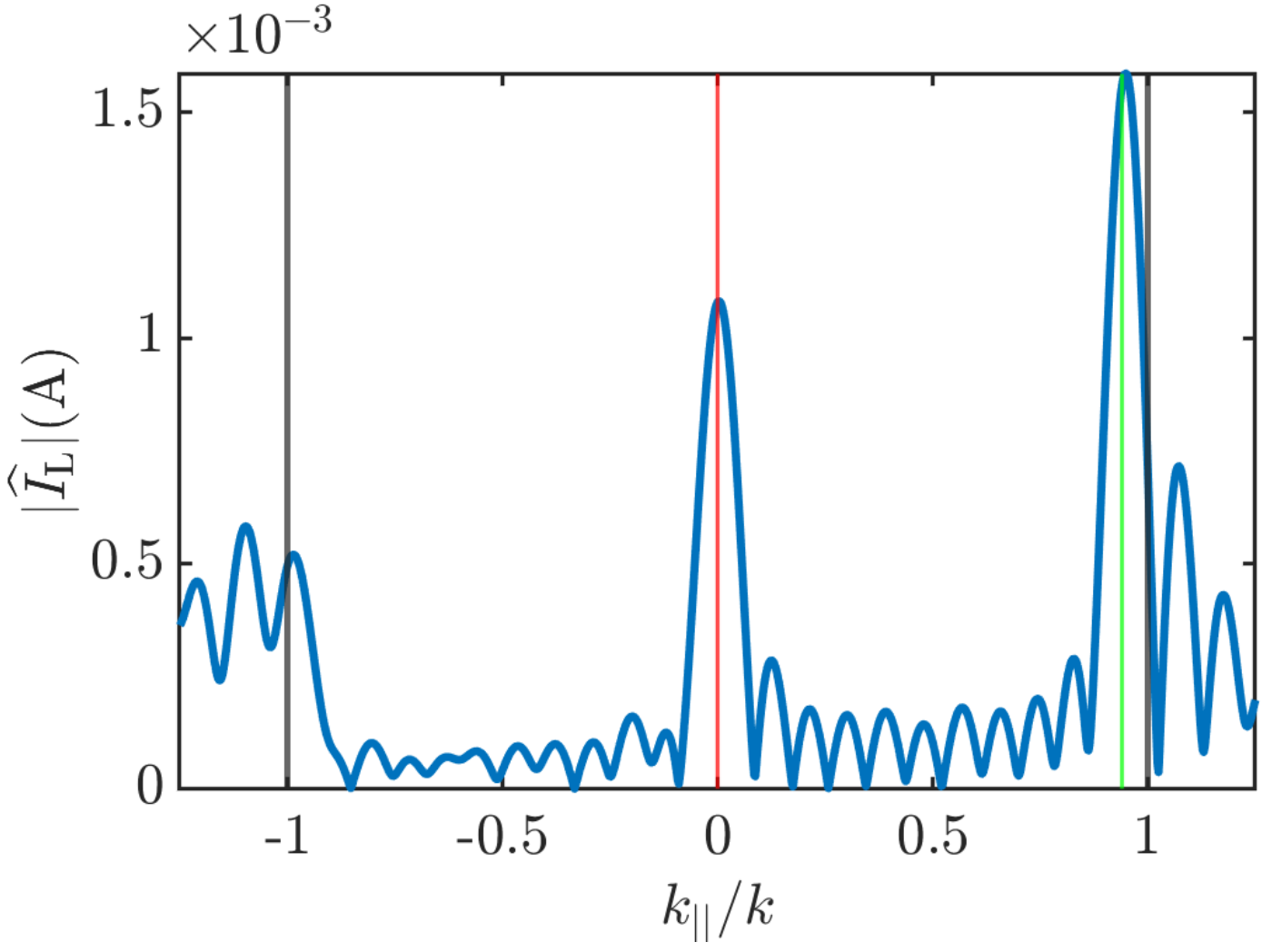}
    \caption{Continuous spatial spectrum of the optimized solution for wire currents when the distance between the wires is set to $d=\lambda/10$. The other key parameters for this solution are: $h=\lambda/6, N=111, r_0=\lambda/100$, and the whole size of the array is $N\times d= 11\lambda$. The frequency is set to  $f=10$~GHz, and the receiver is located at $\theta_ {\rm r}= 70^\circ$.}
    \label{fig:freq_spect_lam_over_ten}
\end{figure}

As was mentioned above, in the optimized case, the total spatial spectrum of the wire currents $\widehat{I}_{\rm {L}} (k_{||})$ will contain another strong spatial harmonic, $\widehat{I}_{\rm {L}}(k_0\sin \theta_{\rm {i}})$, that will be naturally generated to cancel the plane wave specularly reflected from the ground plane of the RIS. The bottom plot in~Fig.~\ref{fig:conceptual}(b) illustrates a qualitative example of the Fourier transform of the current in the optimized case. The highest peak is located at $k_{||}=k_0 \sin \theta_{\rm {r}}$ and the second peak at the specular direction $k_{||}=k_0 \sin \theta_{\rm {i}}$. Since the optimized array exhibits nonlocal response, we will detect some nonzero spatial harmonics at $|k_{||} |>k_0$ that correspond to surface modes and do not contribute noticeably to the radiated power from the RIS. The peak at $k_{||}=k_0 \sin \theta_{\rm {i}}$ reveals the incident wave angle of arrival. Thus, we see that the current optimization method allows us to find $\theta_{\rm {i}}$ without the need of any additional measurements for AoA sensing, but simply observing the spatial spectrum of the induced currents.  Thus, the method provides simultaneous maximization of the signal reception and estimation of the AoA.

\begin{figure}
    \centering
         \includegraphics[width=0.5\textwidth]{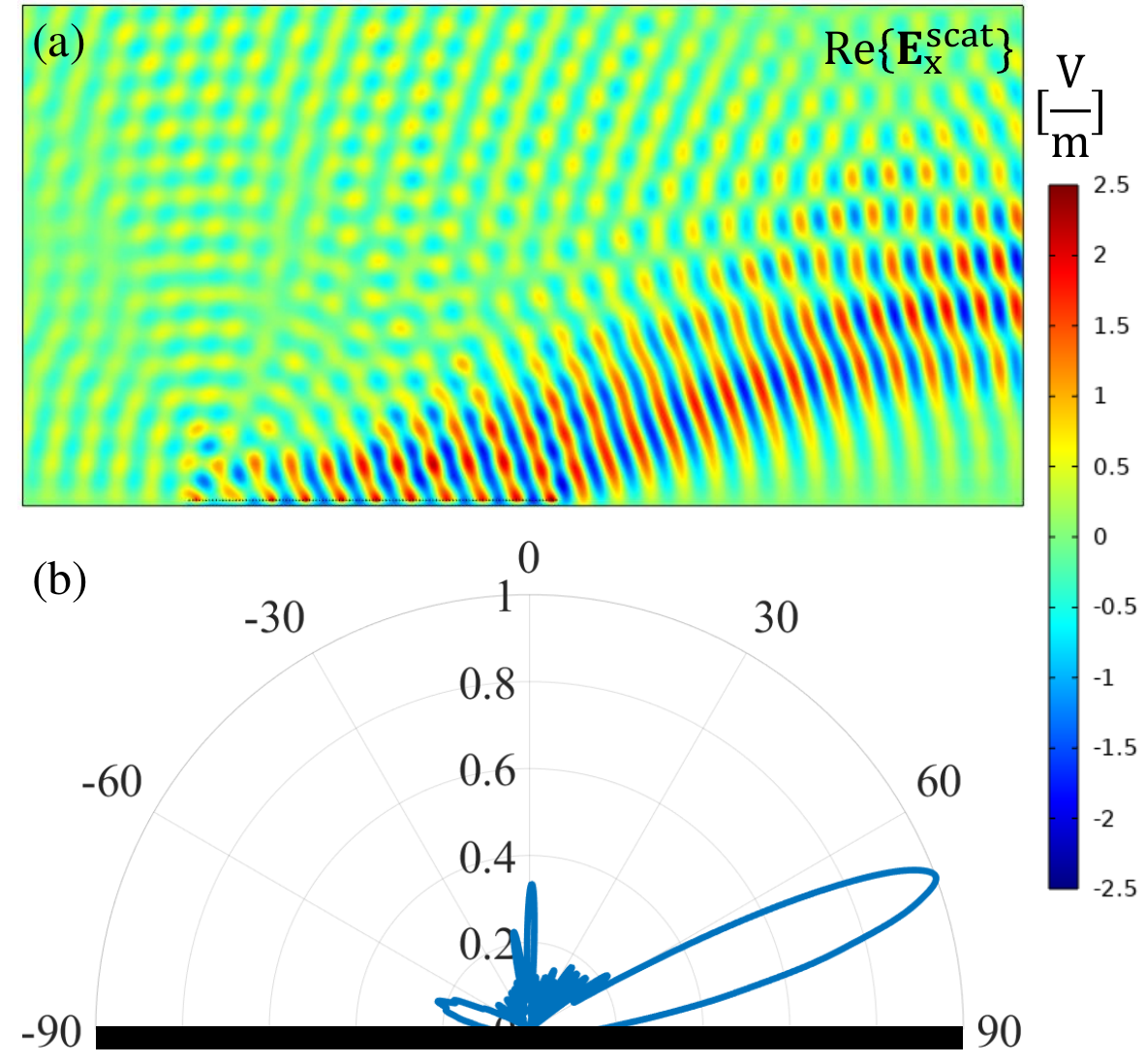}
    \caption{a) The optimized total reflected (sum of the field generated by wires and by the ground plane) field distribution based on the objective function ~Eq.~\eqref{Eq:first_obj}, when there are 111 wires with $\lambda/10$ distance between them. (b) 
The corresponding far-field scattering pattern. In this calculation, we assume a finite-sized ground plane, of the same size as the array of loaded wires.}
    \label{fig:reflected_field}
\end{figure}

The objective function of this optimization method can be defined as follows:
\begin{equation}
\label{Eq:first_obj}
\mathcal{O} = - {\rm min} \left \{ \left| \widehat{I}_{\rm {L}}(k_0\sin \theta_{\rm {r}}) \right| \right \}.
\end{equation}
Optimization is performed in  MATLAB using \textit{fmincon} function, that essentially minimizes the given parameter, which is the reason behind the negative sign in Eq.~\eqref{Eq:first_obj}. This way the current of a certain spatial harmonic is maximized.

A typical optimization result is illustrated in~Fig.~\ref{fig:freq_spect_lam_over_ten}, and the parameters of optimization are shown in the caption. Figure~\ref{fig:freq_spect_lam_over_ten} provides a visualization of the amplitudes of various harmonics resulting from the Fourier transform of the current distribution. The horizontal axis represents the normalized tangential wavenumber along the surface (along the $y$-direction in the considered geometry) equal to the sine of the propagation angle ($k_{||}= k_0\sin \theta$). Additionally, the black vertical lines denote the boundaries of the propagation harmonic range, signifying that within the range limited by these two lines, the tangential wavenumber is smaller than the free-space wavenumber, allowing these harmonics to propagate in free space. Conversely, harmonics falling outside this range correspond to evanescent modes that propagate along the surface and cannot extend to the far zone. 
The red and green vertical lines correspond to the harmonics associated with the direction of the incident and the desired reflected wave, respectively. Notably, since all the load impedances are chosen to be lossless ($ Z^*_{{\rm L},n} =-Z_{{\rm L},n}$), maximizing this current spatial frequency inherently maximizes the efficiency of anomalous reflection by the RIS for any unknown $\theta_{\rm {i}}$. As previously stated, the optimization yields two peaks. The highest peak is aligned closely with the green vertical line indicating that the optimization goal has been achieved and the scattered power is rerouted in that direction. The other peak reveals the direction of the incoming wave ($\theta_{\rm i}=0^\circ$ in this example), as the optimized solution includes a harmonic capable of canceling the specular reflection from the ground plane. Thus, the optimized results depicted in Fig.~\ref{fig:freq_spect_lam_over_ten} substantiate our claims.

To illustrate the optimized scattering, we can calculate the total scattered  electric field by combining the field generated by the wires with the reflected field from the ground plane,
 \begin{equation}\mathbf{E}^{\rm sca} = \mathbf{E}^{\rm wires} + \mathbf{E}^{\rm ref},
 \label{eq:scat}
 \end{equation}
 where
 \begin{subequations}
\begin{align}
   \mathbf{E}^{\rm wires}=&- \frac{k_0 \eta }{4}  \sum_{m=0}^{N-1} 
 I_m \bigg[ H_0^{(2)} \big(k_0 \sqrt{( y - y_m )^2+ ( z+h)^2 } \big)\notag\\
 & -  H_0^{(2)}\big(k_0 \sqrt{( y - y_m )^2+ (z-h)^2 } \big) \bigg] \hat{x}, 
 \label{eq:electric_field_strips}\\
 \mathbf{E}^{\rm ref} &= - E_0 e^{-j k_0 (\sin \theta_{\rm i} y-\cos \theta_{\rm i} z)} \hat{x},
 \label{reflected}
\end{align}
 \end{subequations}
 and $E_0$ is the amplitude of the incident plane wave.
The total reflected field distribution and the radiation pattern for the optimized load values are plotted in~Fig.~\ref{fig:reflected_field}. The results are consistent with what we see in~Fig.~\ref{fig:freq_spect_lam_over_ten}: The total field is scattered predominantly in the desired direction, while the specular reflection from the ground plane is canceled.

Regarding the optimization speed, it is important to note that the input and mutual impedances remain constant. That is, they need to be pre-computed or measured only once. Indeed, only the load reactances are subject to optimization. We remind that in applications of this method, the load currents or voltages are measured, and the only calculation at each optimization step is the trivial summation \eqref{Eq:spatial_spect}.  


\begin{figure}[tb]
     \centering
         \includegraphics[width=0.5\textwidth]{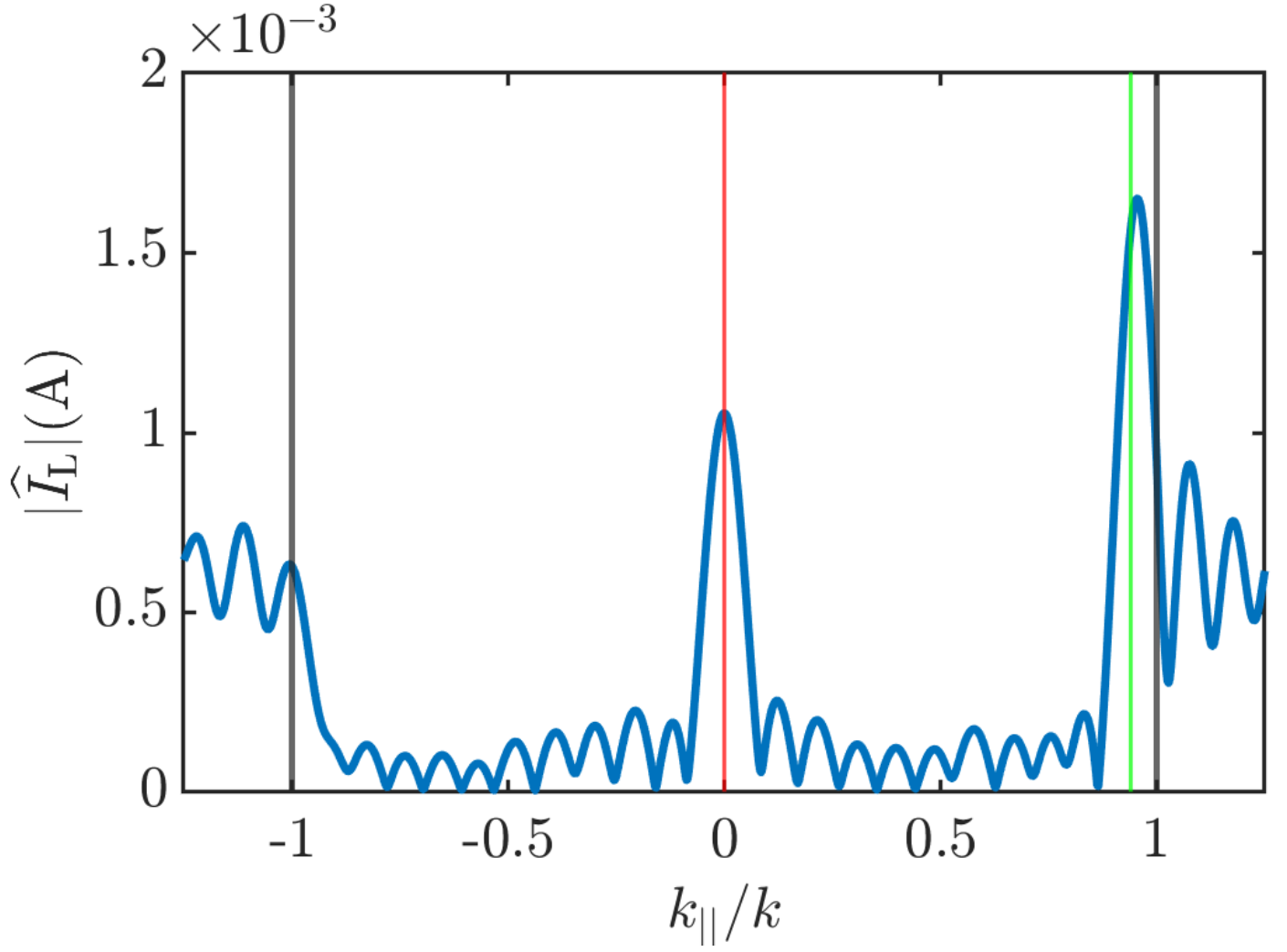}
    \caption{Spatial spectrum of the optimized solution (based on~Eq.~\ref{Eq:far_obj}) for wire currents when the distance between the wires is set to $d=\lambda/10$. The other key parameters for this solution are: $h=\lambda/6, N=111, r_0=\lambda/100$, and the whole size of the array is $N\times d= 11\lambda$. The blue line represents all current harmonics, while the green and red vertical lines correspond to the harmonics associated with the reflected and incident wave directions, respectively. Two black lines delineate the boundaries of the propagation range, where $k_{||}=k_0$.}
    \label{fig:cont_spect_sec_obj}
\end{figure}

\subsection{Optimization based on the electric field in the far zone}

\subsubsection{Single incident wave}
One viable alternative for the objective function is the maximization of the reflected wave amplitude in the direction corresponding to the location of the receiver in the far-zone region:
\begin{equation}
\label{Eq:far_obj}
\mathcal{O} = -{\rm min} \left \{ \left|\mathrm{ \bf E}^{{\rm wires}}(\mathrm{ \bf r},\theta_{\rm {r}},\overline{\overline{Z}}_\mathrm{L}) \right|^2 \right \}.
\end{equation} 
This value can be computed from measured induced load currents and pre-computed effective heights of the array elements or, if there is a data link with the receiver, measured at the receiver position. In the last case, no measurements at the RIS site are needed for optimization of wave reflection. 

\begin{figure}[tb]
     \centering
         \includegraphics[width=0.5\textwidth]{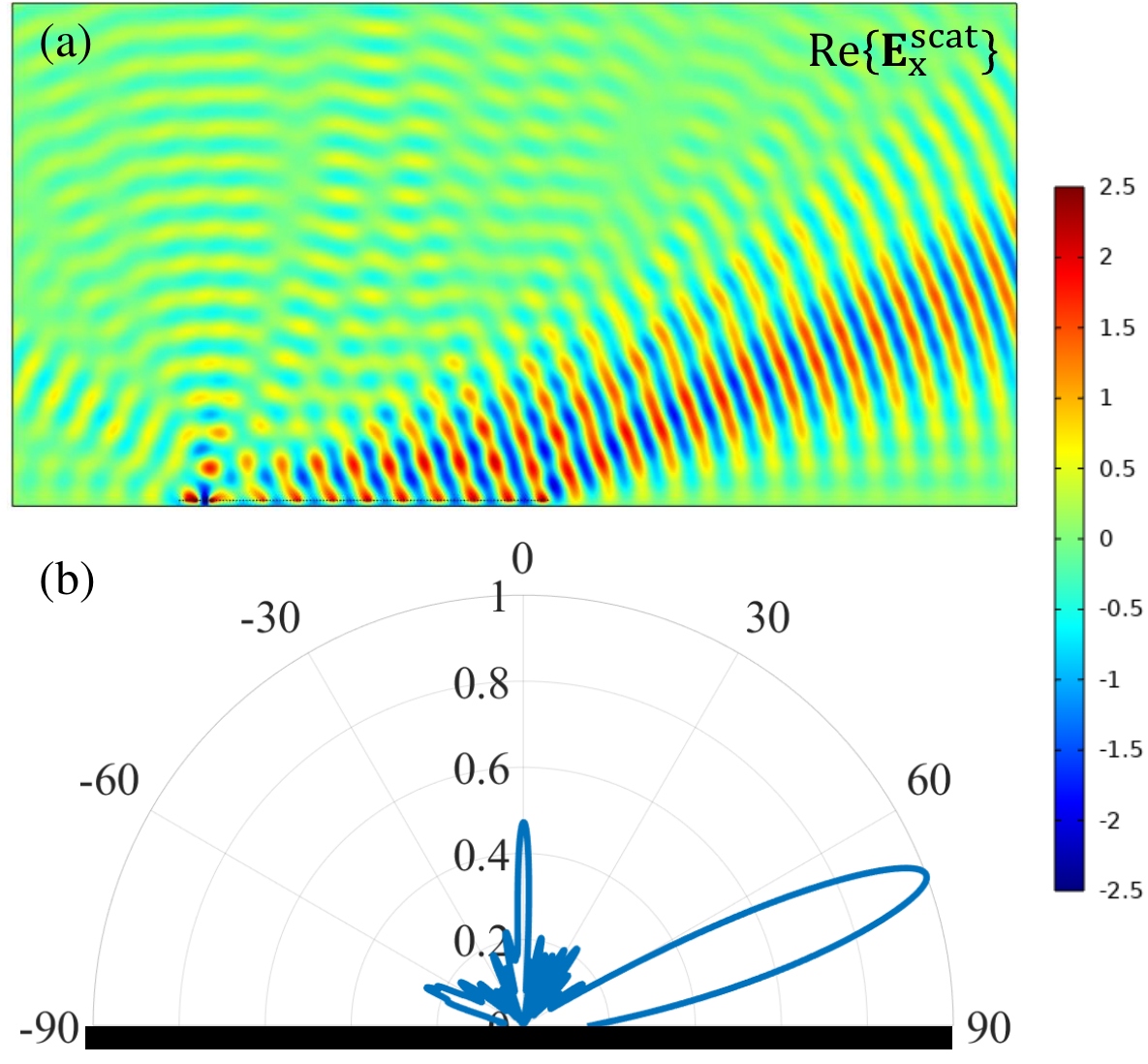}
    \caption{(a) Optimized total reflected field  (sum of the field generated by wires and by the ground plane) distribution based on the objective function as in~Eq.~\eqref{Eq:far_obj}, when there are 111 wires with $\lambda/10$ distance between them. (b) 
The corresponding far-field scattering pattern. In this calculation, we assume a finite-sized ground plane, of the same size as the array of loaded wires.}
    \label{fig:tot_field_seond_appr}
\end{figure}

We perform an investigation employing the same configuration as in the first approach, with 111 wires spaced by $\lambda/10$ at $f=10$~GHz when the receiver direction is fixed to $\theta_{\rm{r}}=70^\circ$. Employing this objective function for optimization, we have found the optimal distribution of load impedances along the array illuminated from the normal direction. 
Subsequently, the final step involves the estimation of the AoA. As in the previous case, this can be achieved by calculation of the spatial spectrum of the optimized current distribution. Figure~\ref{fig:cont_spect_sec_obj} shows the spatial spectrum for this case. The second peak is correctly aligned with the red line (revealing that the incidence angle equals $\theta_{\rm i}=0^\circ$), while the first peak is matched with the green line that shows the direction to  the receiver. 

Figure~\ref{fig:tot_field_seond_appr}(a) shows the total reflected field distribution above the structure. We observe that the field scattered by the wires in the specular direction (which corresponds to the peak at zero $k_{||}$ on Fig.~\ref{fig:cont_spect_sec_obj}) and the directly reflected field from the ground plane cancel each other. Therefore, most of the scattered  power goes to the desired direction, in this example, to $\theta_{\rm r}=70^\circ$. A polar plot of the scattering pattern of the optimized array is depicted in Fig.~\ref{fig:tot_field_seond_appr}(b) and confirms this conclusion. 


\begin{figure}[tb]
     \centering
         \includegraphics[width=0.5\textwidth]{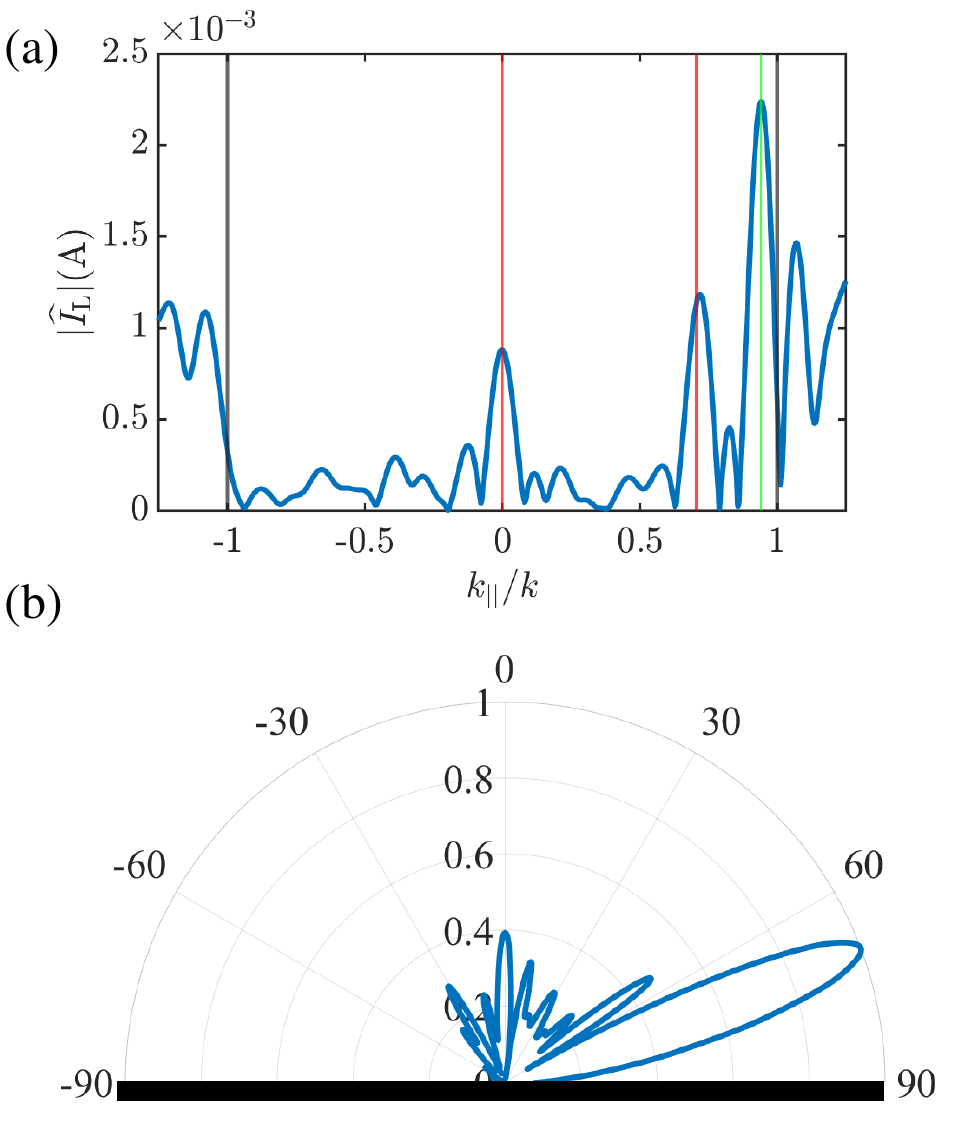}
    \caption{(a) The spatial spectrum for the case $N=111$ wires distanced by $\lambda/10$ and optimized with the objective function defined in~Eq.~\eqref{Eq:first_obj} and the constraint introduced in~Eq.~\eqref{Eq:constraint} when the structure is illuminated by two incident waves form different unknown directions. (b) The scattering pattern in the polar format.}
    \label{fig:multiple_inc}
\end{figure}

Although the above results demonstrate that the AoA can be successfully determined from the optimized current distributions found by both considered optimization approaches, let us briefly compare the performance  of  each optimization method. To do that, we use the efficiency definition in terms of reflected fields in the far zone~\cite[Section~II.B]{li2023all}.
\begin{equation}
\zeta = \frac{\vert \mathrm{\bf E}^{\rm wires}(\theta_{\rm r}, \overline{\overline{Z}}_\mathrm{L}) \vert ^2} {\vert \mathrm{\bf E}_{\rm refer}^{\rm far} (\theta_{\rm r}) \vert ^2}.
\label{eq:efficiency}
\end{equation}
Here, $\vert E^{\rm wires}(\theta_{\rm r}, \overline{\overline{Z}}_\mathrm{L}) \vert$ represents the electric field amplitude produced by wires ($\mathbf{E}^{\rm wires}$ given by \eqref{eq:electric_field_strips}) in the far zone along the direction $\theta_{\rm r}$ for the optimized values for loads  $\vec{Z}_{\rm L}$. Meanwhile, $E_{\rm refer}^{\rm far} (\theta_{\rm r})$ corresponds to the reference far-zone electric field in the same desired direction, generated by an idealized set of current lines of uniform amplitude and linearly varying phase. The current amplitude of this array of radiating wires is adjusted so that an infinitely extended version of it would achieve the perfect power balance of the incident and reflected waves. This efficiency definition is motivated by the fact that, for non-superdirective arrays, a finite-size array with the uniform current distribution attains the highest possible directivity. In cases where the optimized array becomes superdirective, this efficiency measure can exceed unity.

\begin{figure}[tb]
     \centering
         \includegraphics[width=0.5\textwidth]{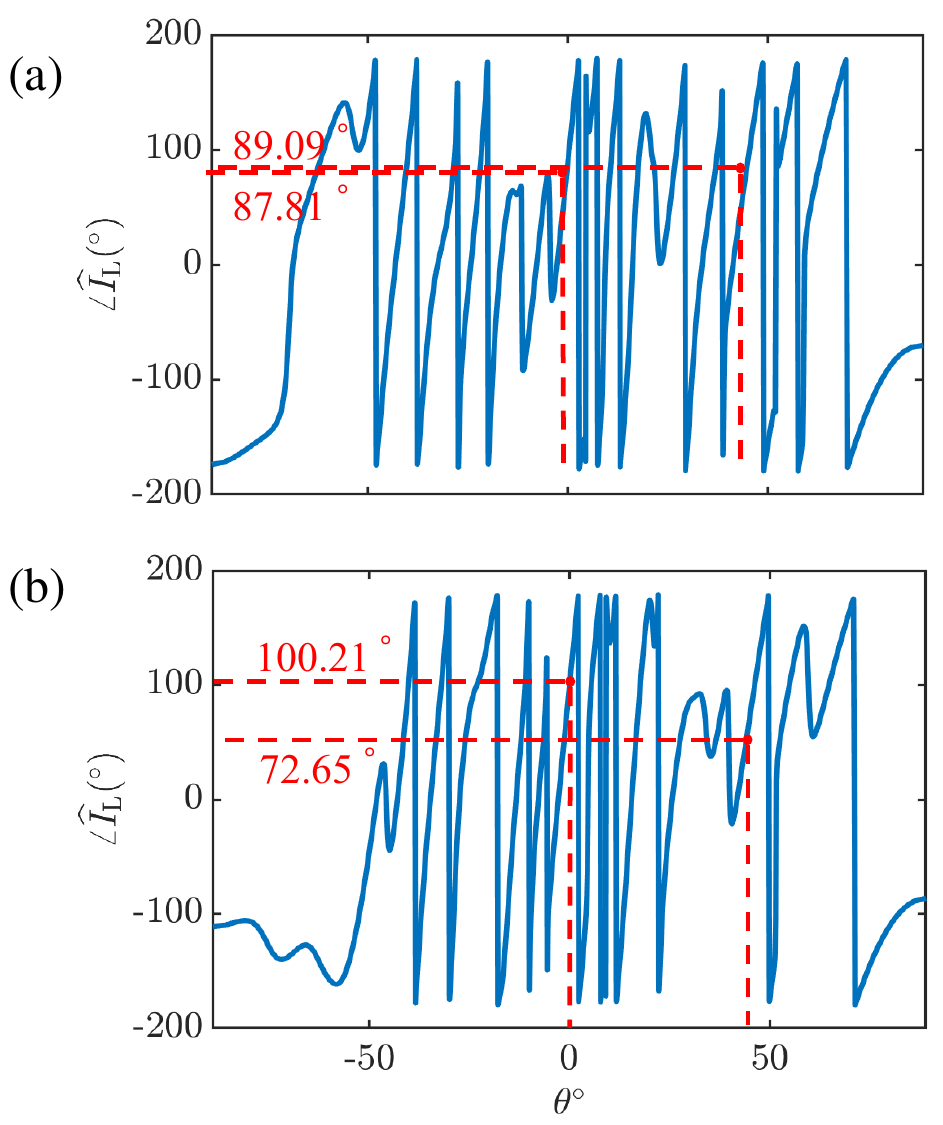}
    \caption{Phase diagram of the spatial spectrum of the current distribution when (a) $\phi=0^\circ$, and (b) $\phi=30^\circ$.}
    \label{fig:phase_digram}
\end{figure}

For a fair and consistent comparison, we adopted an identical configuration consisting of 111 wires spaced at intervals of $\lambda/10$ for both optimization scenarios, each targeting different objective functions. 
When maximizing the current at the desired spatial frequency (the first approach), the maximum efficiency value as defined in~Eq.~\eqref{eq:efficiency} is $\zeta=115\%$ at the reflection angle $\theta_{\rm r}=69.5^\circ$.
However, by utilizing the objective function \eqref{Eq:far_obj} centered on the maximization of the far field exactly at the desired angle, both efficiency and accuracy of the optimized reflection angle show slight improvements, yielding an efficiency of $\zeta=116\%$ at $\theta_{\rm r}=70^\circ$. In essence, this optimization approach results in a more superdirective solution compared to the first case. 


\subsubsection{Multiple incident waves}
In this section, we explore the capability of the proposed method to sense the AoA in scenarios where multiple waves hit the structure from different unknown directions. The investigation focuses on the introduced objective function to maximize the electric field reflected by the array to the desired direction. 
This scenario corresponds to realizing a power combiner for multiple incident waves into a given single reflection direction. It is known that for \textit{coherent} beams, perfect power combination of waves (without spurious reflections) is possible only when the incident beams have specific phase and amplitude relations among them. In other words, once the structure is designed to combine perfectly a given set of incident beams, it will not work ideally anymore once the phases or amplitudes of the incident beams are modified.  
This constraint originates from the passivity of the system~\cite{he2017possibility,hanna2013coherent,wang2021space}.

Let us assume that the structure is illuminated by two incident waves such that the amplitudes are equal and there is no phase difference between them. While the underlying theory and methodology remain consistent with the case of a single incident wave, achieving a more accurate estimation of AoA necessitates introduction of additional optimization constraints.  To this end, the solution is limited to the case when the power scattered to the other angles is minimized. Consequently, we incorporate~Eq.~(\ref{Eq:far_obj}) into our framework, augmented by an extra constraint outlined as follows:
\begin{equation}
\label{Eq:constraint}
\begin{aligned}
&\left|\mathrm{\bf E}^{{\rm wires}} (\theta, \overline{\overline{Z}}_\mathrm{L}) \right| < \delta \\
\end{aligned}
\end{equation}
\begin{equation}
\begin{aligned}
&\theta \in [-90^\circ, \theta_{\rm {r}}-3^\circ], [\theta_{\rm {r}}+3^\circ, +90^\circ], \nonumber
\end{aligned}
\end{equation}
where $\delta$ is a small positive constant that quantifies the reradiated power that is allowed toward the specified angular sectors.
Therefore, the optimizer searches for a solution that maximizes reflected power into the desired direction while simultaneously minimizing power in other directions, except for a narrow beamwidth (in this case, we have chosen $\pm 3^\circ$). By incorporating this additional constraint into the optimization setting, using the same parameters as the single-illumination scenario discussed earlier, we successfully perform the optimization, and the results are presented in~Fig.~\ref{fig:multiple_inc}. Figure~\ref{fig:multiple_inc}(a) depicts the continuous spatial spectrum of currents in the array, revealing three peaks. The highest peak corresponds to the desired direction (indicated by the green line along $70^\circ$), while the other two indicate incident waves originating from $\theta_{\rm i} = 0^\circ$ and $-\theta_{\rm i} = 45^\circ$ (shown by red vertical lines). Figure~\ref{fig:multiple_inc}(b) displays the radiation pattern for the optimized case. As anticipated, the majority of the power is directed toward the desired direction, while the reflected power in other directions, excluding the narrow beamwidth, is significantly lower than the main beam. 

This result validates the effectiveness of the proposed method for more practical scenarios where multiple incident waves are impinging upon the structure. In addition, we note that the proposed approach allows us to determine not only the directions of arrival of several simultaneous incident waves, but also the phase differences between these excitations. Indeed, the phase difference of the incident waves is approximately equal to the phase difference between the corresponding peaks on the spatial spectrum of the optimized current distribution. To test this functionality, we assign  an arbitrary  phase difference $\phi$ between the electric fields of two plane waves illuminating at different angles: 
\begin{equation}
\angle \mathbf{E}_{\rm {inc}1}-\angle \mathbf{E}_{\rm {inc}2} = \phi
\label{eq:phase_diff}
\end{equation}
and perform optimization of the induced currents or far fields. For instance, for $\phi=0^\circ$ and $\phi=30^\circ$,  phase diagrams of the spatial spectrum of the optimized current distribution are plotted in~Fig.~\ref{fig:phase_digram}. For $\phi=0^\circ$, the detected phase difference is $1.28^\circ$. When the actual phase difference is $30^\circ$, the corresponding retrieved value is $27.6^\circ$. 
These small deviations are due to the fact that the two coherent beams were reflected to the desired direction at $\theta_{\rm r}$ with the efficiency slightly smaller than the ideal one. Indeed, as is seen from Fig.~\ref{fig:multiple_inc}(b), there are sidelobes centered at  $\theta_{\rm i}=0^\circ$ and  $-\theta_{\rm i}=45^\circ$, showing that the specular reflections from the ground plane are not completely compensated, which yields the aforementioned slight phase error.

\section{Conclusion}

We have presented a general concept of dual-functional RIS, where the measured induced voltages or currents at the controllable loads are used to tune reflection of the incident radiation into the desired direction and, at the same time, to determine the unknown angles of arrival of the incident radiation. 
We expect that  integration of sensing capabilities into tunable anomalous reflectors offers a compelling prospect for the future of mobile networks. Comparing with known alternative solutions, the most important advantage of the proposed concept is that there is no need for any  additional sensing components and measurement electronics.  As a result, it becomes possible to  implement thin, lightweight, and nearly passive RIS arrays with this double functionality. 

To validate and illustrate our concept, we have used an example of a simple array of loaded wires, in which case it is possible to study it using fully analytical means.   Importantly, the introduced concept is applicable to arrays with arbitrary shapes of the array elements, as long as the measurement of the induced current or voltage distribution within that structure is feasible. 

We have used two distinct objective functions aimed at optimizing reflections from  aperiodically loaded arrays (from an electromagnetic perspective) in real-time scenarios. Our comparative analysis of optimizations based on these objective functions in terms of accuracy and efficiency showcased their potential. Furthermore, by considering the case of multiple simultaneous illuminations, we have shown that the introduced method is applicable also in this more general case, performing sensing of multiple angles of arrival. Our future endeavors will focus on applying the theoretical framework presented here to practical applications. We think that this work can be useful for enhancing communication systems, particularly for continuous AoA tracking of users of mobile networks.

\section*{Acknowledgment}
This work was supported by the European Union’s Horizon 2020 research and innovation programme under the Marie Skłodowska-Curie grant agreement No 956256 (project METAWIRELESS) and by the Finnish Research Council (Academy of Finland), grant 345178. Y.L. acknowledges support from the China Scholarship Council under Grant 202106280229. G.P. acknowledges support from Ulla Tuominen Foundation.

\nocite{*}
\bibliographystyle{IEEEtran}
\bibliography{apssamp}
\end{document}